\documentclass{article}
\usepackage[dvips]{epsfig}
\usepackage{cite,amsmath,amssymb,dcolumn}
\begin{document}
\title{Automatic Coarse Graining of Polymers}
\author{Roland Faller\footnote{Email: rfaller@ucdavis.edu} \\
\small Department of Chemical Engineering \& Materials Science,\\
\small University of California-Davis, Davis, CA 95616, USA}
\maketitle
\abstract{
\noindent Several recently proposed semi--automatic and
fully--automatic coarse--graining schemes for polymer simulations 
are discussed. All these techniques derive effective potentials 
for multi--atom units or super--atoms from atomistic simulations. 
These include techniques relying on single chain simulations in vacuum and
self--consistent optimizations from the melt like the simplex
method and the inverted Boltzmann method. The focus is on matching
the polymer structure on different scales. Several ways to obtain
a time-scale for dynamic mapping are discussed additionally.
Finally, similarities to other simulation areas where automatic
optimization are applied as well are pointed out.}

\section{Introduction}
Polymers with their large variety of important length scales pose a
formidable challenge for computer simulations. Over the last decades
various techniques to handle the problems on the different length
scales separately have been developed. Especially simulations in full
atomistic
detail~\cite{paul95,moe96b,mplathe96a,antoniadis98,faller01a}, and
with one interaction center for each 
monomer~\cite{grest86,kremer90,faller00b} or for each
polymer~\cite{murat98} have gained a lot of attention.

More recently it has been realized that a connection between the
arising length and time scales are necessary. To this end a number
of coarse--graining techniques have been
devised~\cite{tschoep98a,mccoy98,baschnagel00a,akkermans01,akkermans00,fukunaga02,mplathe02,reith03,faller03a,mplathe03,tsige03} 
where simulations on more 
than one length scale are combined in order to get a better
understanding of the system as a whole.  It has even been proposed
that simulations on both scales can be performed in one single
simulation box~\cite{mccoy98,tsige03}. The purpose of this contribution is
to critically analyze several of the most recent automatic mapping
schemes for coarse--graining in polymer research. This comprises a
technique combining atomistic single chain Monte Carlo with
molecular dynamics on the meso--scale~\cite{tschoep98a} the
automatic simplex mapping technique~\cite{meyer00,reith01a}, and
a number of physically inspired 
techniques~\cite{akkermans01,faller03a,reith03}. All these techniques
have been implemented in automatized schemes which in principle
allow to obtain a coarse--grained polymer model on the meso--scale
without human intervention if the atomistic simulations have been
performed. Based on the atomistic simulations a target function has to
be defined and optimized against in the meso--scale simulation.

Techniques which either rely on the use of lattice simulations or which 
cannot be implemented in an automatic manner have been left out on
purpose in this contribution. The reader is referred to other
reviews including such techniques~\cite{baschnagel00a,mplathe02,mplathe03}.

There are many reasons for applying coarse--graining schemes for
polymer simulations. The overall structure of a polymer in melt or
solution often shall be reproduced faithfully except for the local
atomistic detail. This improves the speed and memory requirements of
the simulation and by that allows larger simulations or longer
chains. Simulations of long chains are necessary but the
experimentally relevant chain lengths cannot be reached by
atomistically detailed simulations. Even if computer speed increases
in the future as it did over the last decades, we are still decades
away from doing simulations of chains with thousands monomers in a
fully atomistically detailed simulated melt. The relevant relaxation
times increase by an exponent of $N^{3.4}$ with chain length $N$ for
large chains~\cite{strobl97}. And, even if it were possible to perform
such simulations their usefulness would be questionable as the vast
amount of data would be very difficult to analyze as the interesting
observables would be difficult to filter out. A lot of questions on
large scales have been answered by simple bead--spring models. These
models are able to get interesting scaling behaviors and by this a lot
of basic understanding. In order to get compare directly to
experiments, however, one needs a meso--scale model which does not
represent generically ``a polymer'' but has an identity of a specific
polymer. To this end a combination of atomistic and meso--scale models
which can be mapped uniquely onto each other is necessary. In this
case issues appearing on different length scales can be answered
consistently.

The remainder of this article is organized as follows. Section 2
deals with the various possibilities of static mapping, section 3
with dynamic mapping which has gained much less and in the end
conclusions will be drawn and connections to other automatic
optimization techniques in molecular simulation will be shown.
\section{Static Mapping}
\subsection{The concept of super--atoms}
The methods to be discussed here deal all with two length scales.
Most often these are the atomistic scale and the meso--scale. However,
for the methods to work this is not necessary. For the remainder of
this contribution an atomistic simulation is defined to be a
simulation where all atoms are present or only the hydrogens are
neglected. The latter is often called a united atom model. A
meso--scale model is defined to be a model where a group of atoms is
replaced by one interaction center. This group is typically of the
size of a monomer. We call such a unit a super--atom.

\begin{figure}
\includegraphics[width=0.9\textwidth]{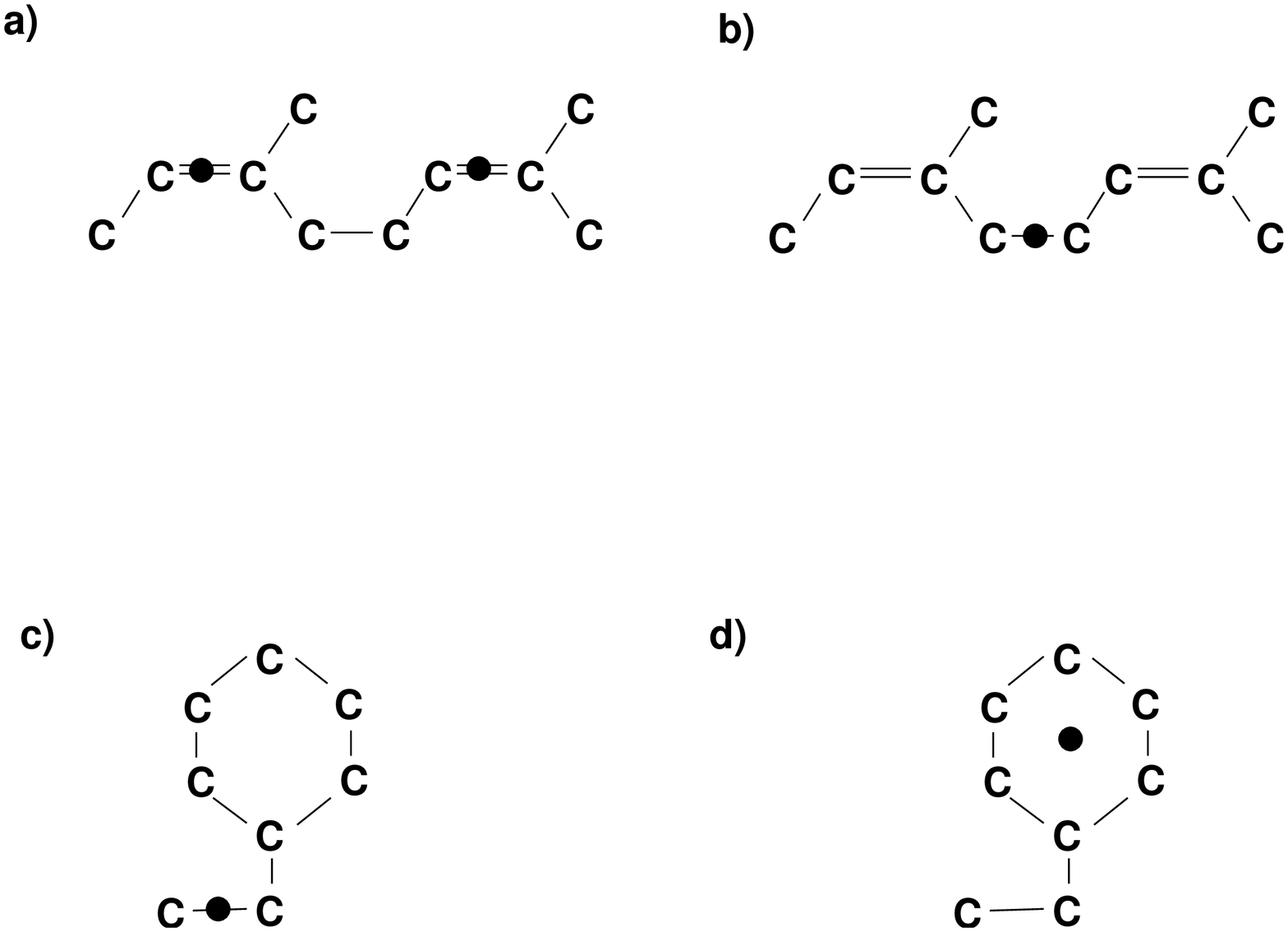}
\caption{Illustration of super--atoms representing
polymers. The hydrogens are left out for clarity. a) {\it
Cis}--polyisoprene physical monomer, center of the super atom in the
middle of the double bond b) polyisoprene pseudomonomer, center of the
super atom between two physical monomers c) Polystyrene. Super--atom
center on the single bond in the monomer d)Polystyrene. Super--atom
center in the center of the ring. All the super--atom centers are
marked by black dots.}
\label{fig:super}
\end{figure}

Thus, a part of a polymer chain comprising a few atoms
(typically 10--30) will be represented by one interaction center. The
super--atoms are the only interaction centers in the meso--scale
simulation. The interaction between super--atoms has to
implicitly carry the information of the interactions between the atoms
in their local geometrical arrangements imposed by the
bonding. Figure~\ref{fig:super} shows some typical examples of super
atom representation of polymers.

\begin{figure}
\includegraphics[width=6cm]{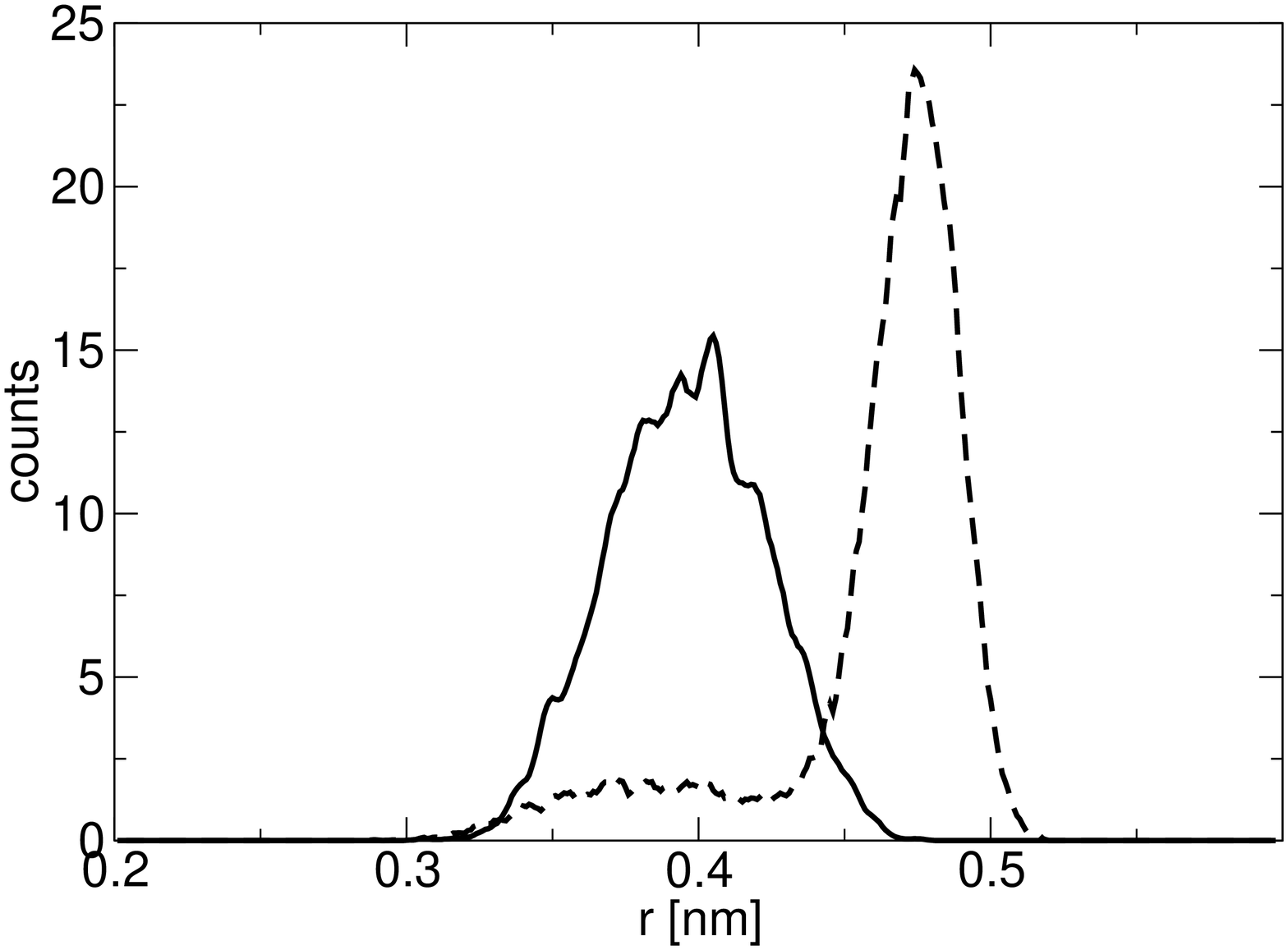}
\includegraphics[width=6cm]{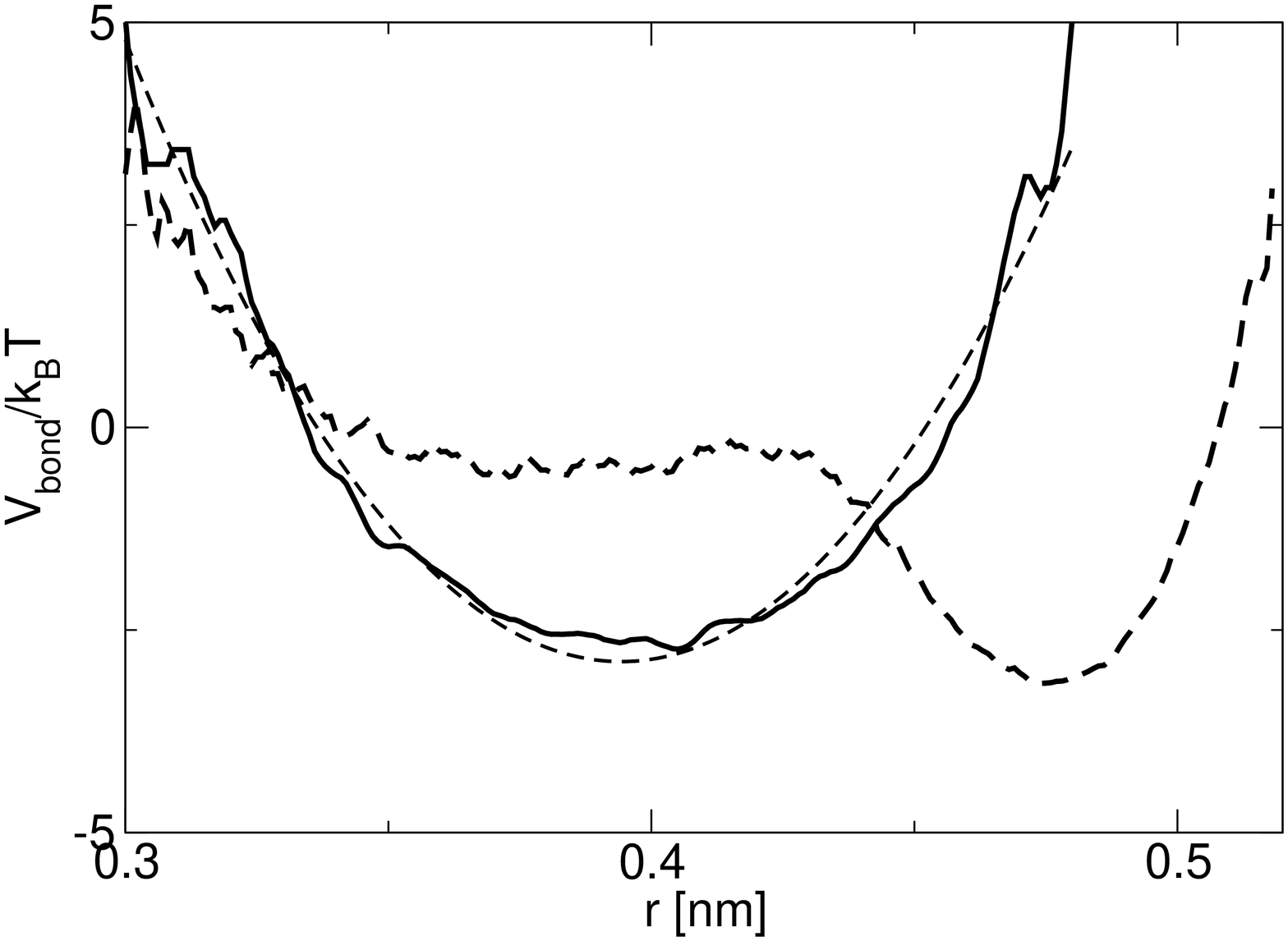}
\caption{ Left: Bond length distributions arising from the possible
choices of super--atoms in {\it cis}--polyisoprene of
Figure~\ref{fig:super}. The single peaked solid line corresponds to
the center of the super--atom on the single bond between atomistic
monomers (Fig.~\ref{fig:super}b), the dashed line to the super--atom
in the center of the double bond (Fig.~\ref{fig:super}a).  All
histograms are normalized that the integral equals 1.  Right: Bond
potentials gained by direct Boltzmann inversion of the distributions
of the left hand side (same line styles).  The thin broken line is a
harmonic fit to the pseudomonomer potential. Curves were locally
smoothed for differentiability.} \label{fig:superchoice}
\end{figure}

The choice of super--atoms is arbitrary in principal. But there are a
number of criteria which have been established. It is very effective
if the distance between super--atoms along the chain is relatively
rigidly defined as in that case the bonding potential is just a
harmonic bond~\cite{faller03a}. Figure~\ref{fig:superchoice}
illustrates this using {\it cis}--1--4--poly--isoprene.  The obvious
choice of the center of mass of the double bond leads to a doubly
peaked bond distribution whereas the choice of the super--atom center
being placed between atomistic monomers results in a clear single
peak. Such a distribution can easily be modeled by a single Gaussian
which is produced by a harmonic bond potential The height to width
ratio of the Gaussian peak defines the harmonic bond strength. The
underlying reason for the two strongly different distributions is that
the double bond is very rigid and does not allow any torsional degrees
of freedom whereas the single bonds can easily flip from one torsional
state to another. As the centers of mass of the double bond are
effectively connected by single bonds and vice versa the double bonds
lead to the sharply peaked distribution and the different torsion
states of the single bonds lead to more than one peak making it more
difficult to model the distribution by a simple bond
potential. Additionally the multiplicity of peaks would lead to an
interdependence of bond and angle potentials.

Moreover, it is advantageous to have the space occupied by the atoms
represented in one super--atom being spherical in order to avoid anisotropic
potentials. Almost all schemes use a spherical potential to model the
space occupied by the
super--atom~\cite{mccoy98,tschoep98a,akkermans01,faller03a,meyer00}.
This occupied space can in the first approximation be viewed as the elliptic
hull of the atoms. Generalizations to anisotropic potentials have been
attempted~\cite{hahn01}. The offset by the much higher complexity of
the simulation can in most cases not been overcome by the only
slightly higher accuracy.

If a single spherical potential is not satisfactory as, e.g. for
diphenylcarbonate~\cite{meyer00} or polycarbonates~\cite{tschoep98a}
it is more economical to use more than one spherical super--atom per
monomer than a non--spherical one. Abrams et al. showed that in the
case of polycarbonate one needs 5 spherical interaction center per
atomistic monomer in order to get a good representation of the
underlying polymer~\cite{abrams03} whereas 3 anisotropic beads have been used by
Hahn et al~\cite{hahn01}.
\subsection{Single chain distribution potentials}
Tsch{\"op} et al proposed a technique for mapping the
structure of a polymer to a model containing much less interaction
sites~\cite{tschoep98a}. The model starts out with a detailed
quantum chemical calculation of short segments of the polymer chain
in order to obtain a accurate torsion potential. This
quantum chemically determined distributions are then used to
perform single chain Monte Carlo simulations in vacuum. The
corresponding distributions of super--atoms are recorded. The
recorded distributions are bond lengths, bond angles and torsions.
In order to accurately gain a potential out of these distributions
they have to be weighted by the corresponding Jacobians. E.g. for
the bond lengths the Jacobian is just $r^2$ which stems from the
transformation from spherical to Cartesian coordinates. Then they
are Boltzmann--inverted to obtain intra--molecular potentials
between super atoms, i.e. a potential is derived from the
distribution. Formally the Boltzmann inversion leads to a free
energy difference but in vacuum this equals the potential energy.
This difference will become crucial in the following sections.
\begin{equation}\label{eq:Boltzmann}
V(\zeta)=-k_BT\ln p(\zeta)
\end{equation}
Here $\zeta$ can stand for bond lengths, bond angles and torsions
alike. The distribution $p(\zeta)$ is taken after the Jacobian
correction. In this way a complete set of intra--molecular potentials
has been obtained. It is noteworthy that this potential is completely
numerical. In order to be able to calculate a derivative to obtain the
forces local splines or similar techniques can be used to smooth it.
Cross dependencies of the different potentials (e.g. bond and angle) 
are neglected for computational reasons. As explained above they can be eliminated 
by the proper choice of mapping points.

In the original work, this elaborate intra--molecular potential was
combined with  a simple repulsive Lennard Jones or  WCA
potential~\cite{weeks71} to reproduce the density. 
\begin{equation}
V_{WCA}=4\epsilon\left[\left(\frac{\sigma}{r}\right)^{12}-\left(\frac{\sigma}{r}\right)^6\right]\quad r<\sqrt[6]{2\sigma}.
\end{equation} 
Here $\sigma$ is the interaction radius (size) of the monomer,
$\epsilon$ is the interaction strength, and $r$ is the distance
between corresponding monomers.

\begin{figure}
\includegraphics[width=0.9\textwidth]{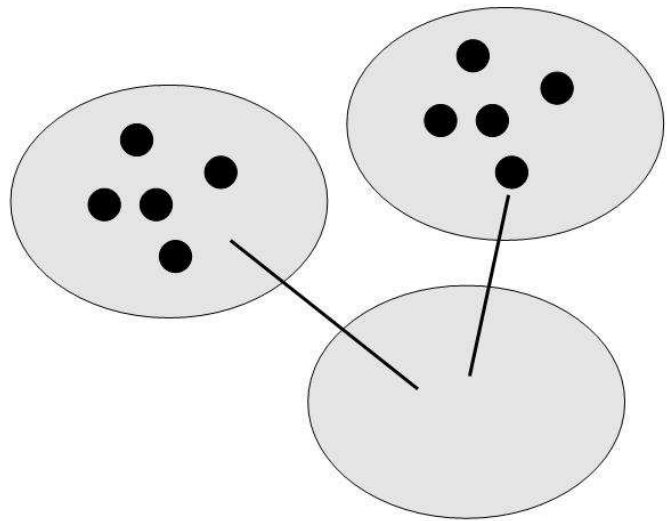}
\caption{The scheme of Mc. Coy et al. uses different degrees of detail
in the very same simulation. The figure shows two particles which
exist on both scales. These interact by their atomistic
potentials. The atomistically detailed interact with the purely
mesoscopic by the mesoscopic potential as do the purely mesoscopic
among themselves. Some of the particles are bonded to form a polymer.} \label{fig:doublepot}
\end{figure}

This method was successful in calculating the structure factor of
polycarbonates~\cite{eilhard99}.  A similar approach has been applied
to a simple hydrocarbon chain where the super--atom center is taken as
the center of mass of $n$ monomers~\cite{fukunaga02}. In this case the
starting point was a atomistic molecular dynamics simulation. For the
non--bonded potential also a potential of mean force approach has been
taken. So the radial distribution function of two dilute polymers is
used to determine the non--bonded potential. For small molecules this
can be used directly~\cite{mccoy98}. For polymers at small distances
the connectivity leads to a severe restriction on the possible
conformations so a restricted pair distribution function taking
connectivity into account has to be used.
This approach does not separate the simulations of the atomistic
and the coarse--grained models and bases on the reversible work
theorem~\cite{mccoy98,tsige03}
\begin{equation}
e^{-\beta W(r)}=\frac{\sum_ie^{-\beta U_i(r)}}{\sum_ie^{-\beta
U_i(\infty)}}
\end{equation}
where $W$ is the reversible work and this can be used as a potential
to obtain the same structure as the atomistic model. The appearance of
the inverse temperature $\beta=(k_BT)^{-1}$ scaled by the Boltzmann
factor $k_B$ makes it clear that this is valid only at the specified
temperature. The potential $U$ is the full potential energy of the
system with the two sites under focus fixed at a distance $r$ apart.
Fully detailed and mesoscopically modeled particles coexist in the
very same simulation. The detailed particles carry two potentials as
they interact with the non--detailed particles as if they were
non--detailed particles (cf. Fig.~\ref{fig:doublepot}). Actually the
two types of particles can even be bonded to each other in order to
get the correct potential along a polymer chain as pointed out in
reference~\cite{fukunaga02} where also the automatic implementation
was shown.
\subsection{Simplex}
Recently more direct ways of linking atomistic melt simulations
and meso--scale melt simulations have been developed. The idea is to
systematically and self--consistently reproduce structure and
thermodynamics of the atomistic simulation on the meso--scale. As
this is an optimization problem mathematical optimization
techniques can be applied directly. One of the most robust although not
very efficient multi--dimensional optimizers is the simplex~\cite{press92}. 
It has the advantage that it does not rely on any
derivatives as they are very difficult to obtain in the
simulation. The simplex was first applied to optimizing atomistic
simulation models to experimental data~\cite{faller99c}. The idea
is to view the experimental observables, e.g. the density $\rho$, as a
function $f$ of the parameters of the simulation model $B_i$, e.g. the
Lennard Jones parameters
\begin{equation}
\rho=f(\{B_i\}).
\end{equation}
This function in multi--dimensional space is now optimized by the
simplex technique. In order for the simplex to be applicable a single
valued function with a minimum at the target has to be defined. This
is easily accomplished by the sum of square deviations from target
values
\begin{equation}
f=\sum_i[A(\{\epsilon_i\},\{\sigma_i\})-A_{\text{target}}]^2.
\end{equation}
Here $A$ represents any thermodynamic observable to be reproduced in this 
scheme with a target value $A_{\text{target}}$; 
$\{\epsilon_i\},\{\sigma_i\}$ are the full set of Lennard--Jones parameters.
Every function evaluation includes a complete equilibration sequence
for the given parameters, a production run and the analysis. In order
to ensure equilibration it was made certain that no drift in the
observables remained and an automatic detection of equilibration was
developed~\cite{faller99c}. Very recently it has been shown that the
derivatives of the observables with respect to the parameters of the
simulation model can also be calculated and therefore more efficient
optimizers can be used~\cite{bourasseau03}.

In the context of polymer mapping the target functions are not
experimental observables but the structure of the system. So radial
distribution functions are the aim of the technique. To this end one
views any point of the radial distribution function $g(R)$ in the
interval $[R_i,R_i+1]$ as a different observable which is to be
reproduced. The function to be minimized is the integral over the
squared difference in radial distribution functions~\cite{meyer00}. If
necessary a weighting function can additionally be
introduced~\cite{meyer00,faller03a}. An exponentially decay is a good
choice as the local structure around the first peak in the rdf is most
crucial and most difficult to reproduce.
\begin{equation}
f=\int \text{d}r w(r)[g(r)-g_{target}(r)]^2.
\end{equation}
A drawback of the simplex technique is that it cannot use
numerical potentials as a relatively small set of
parameters defining the parameter space is needed. The limit is typically
4--6 independent parameters $B_i$. An increase in dimensionality of
this space increases the need for computational resources
tremendously. A good choice for such parameters are a
Lennard--Jones like expansion~\cite{meyer00,reith01a}
\begin{equation}
V(R)=\sum_i\frac{B_i}{r^i}
\end{equation}
where $i$ has been used to span the even numbers from 6 to 12. This
technique has been successful to reproduce monomers of
polyisoprene~\cite{meyer00}. The structure of small molecules like
diphenylcarbonate could be described by this technique as
well~\cite{meyer00}. The application to polymers showed some
deficiencies~\cite{reith01a} which led to the development of better suited
algorithms.
\subsection{Physically Inspired Optimization Methods}
The Iterative Boltzmann method was developed in order to circumvent
the problems encountered with the simplex
technique~\cite{reith03,faller03a}. It is an optimization aiming at
the structure of an atomistic simulation.  It showed its strength by
being able to reproduce the structure of {\it trans}--1,4--polyisoprene
where the simplex technique failed~\cite{reith03,faller03a}. The
idea is to use a physically inspired optimization technique to speed
up the convergence and at the same time get rid of the limitation on
the number of parameters as imposed by the simplex technique.

As discussed above, in the limit of infinite dilution one could use
the potential of mean force gained by Boltzmann inverting the pair
distribution function to get an interaction potential between
monomers, this would be the non--bonded generalization of the above
described single chain approach. Similar ideas have been used to
calculate potentials of mean force (PMF) of large particles like
colloids in matrices of small particles where the small particles play
only the role of a homogeneous background~\cite{engkvist96,kim02}. In
concentrated solutions or melts the structure is defined by an
interplay of the PMF and the packing of atoms or monomers. It has been
shown that simple packing arguments can account for the largest part
of local orientation correlations in dense
melts~\cite{mplathe00}. Thus, a direct calculation of the potential of
mean force is not correct. Still the use of the PMF idea as a way to
iteratively approach the correct potential is possible and is used by
the iterative Boltzmann method. A melt or solution of polymers is
simulated in atomistic detail to obtain a pair distribution
function. For every iteration a one--to--one correspondence between
the effects at a distance $r_0$ and the potential $V(r_0)$ (or force
$-\text{d}_rV(r)|_{r=r_0}$) at the same distance $r$ is assumed.
However, this is not a limitation as the iterative procedure takes 
care of any other dependencies.

\begin{figure}
\includegraphics[width=0.8\textwidth]{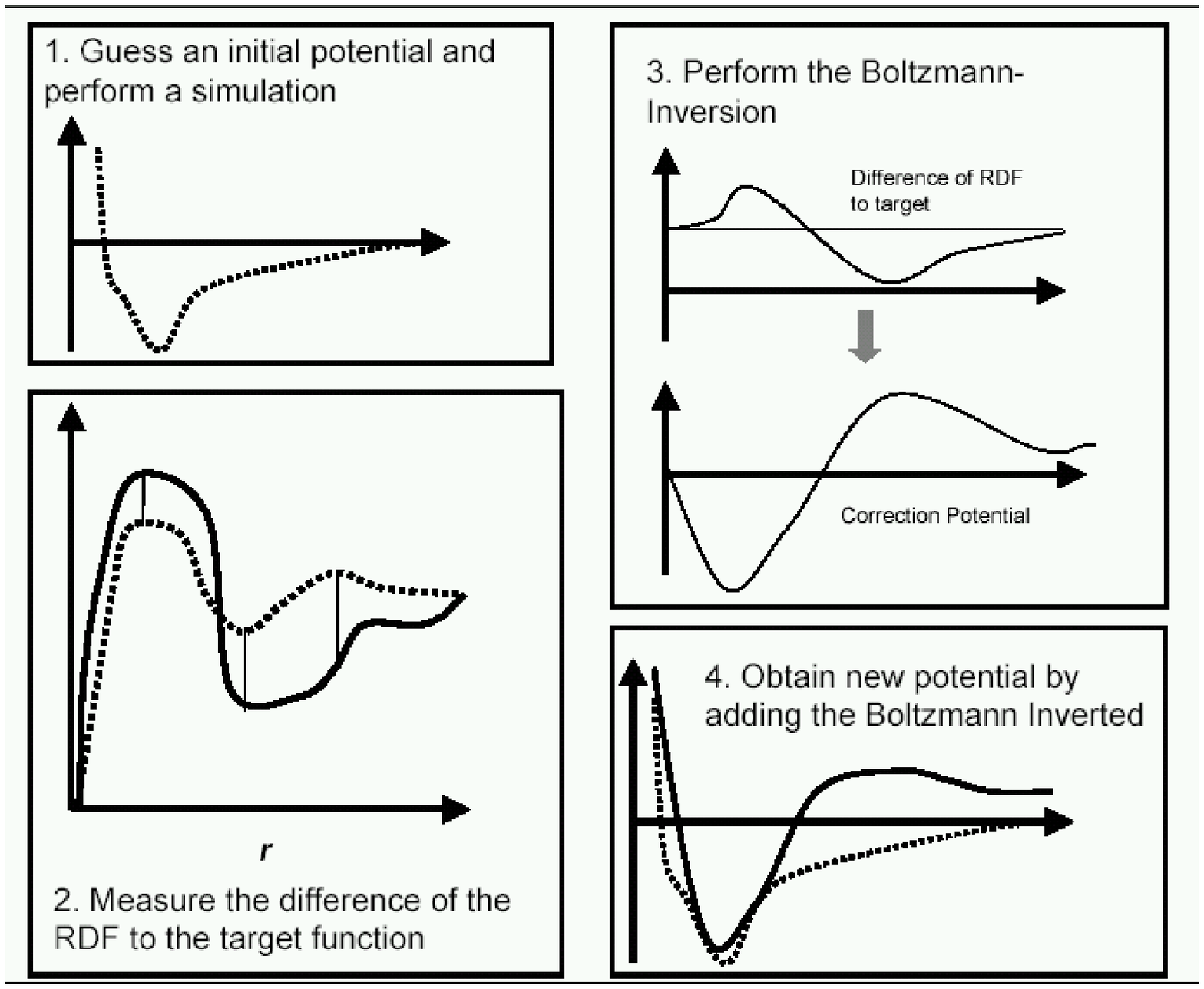}
\includegraphics[width=0.8\textwidth]{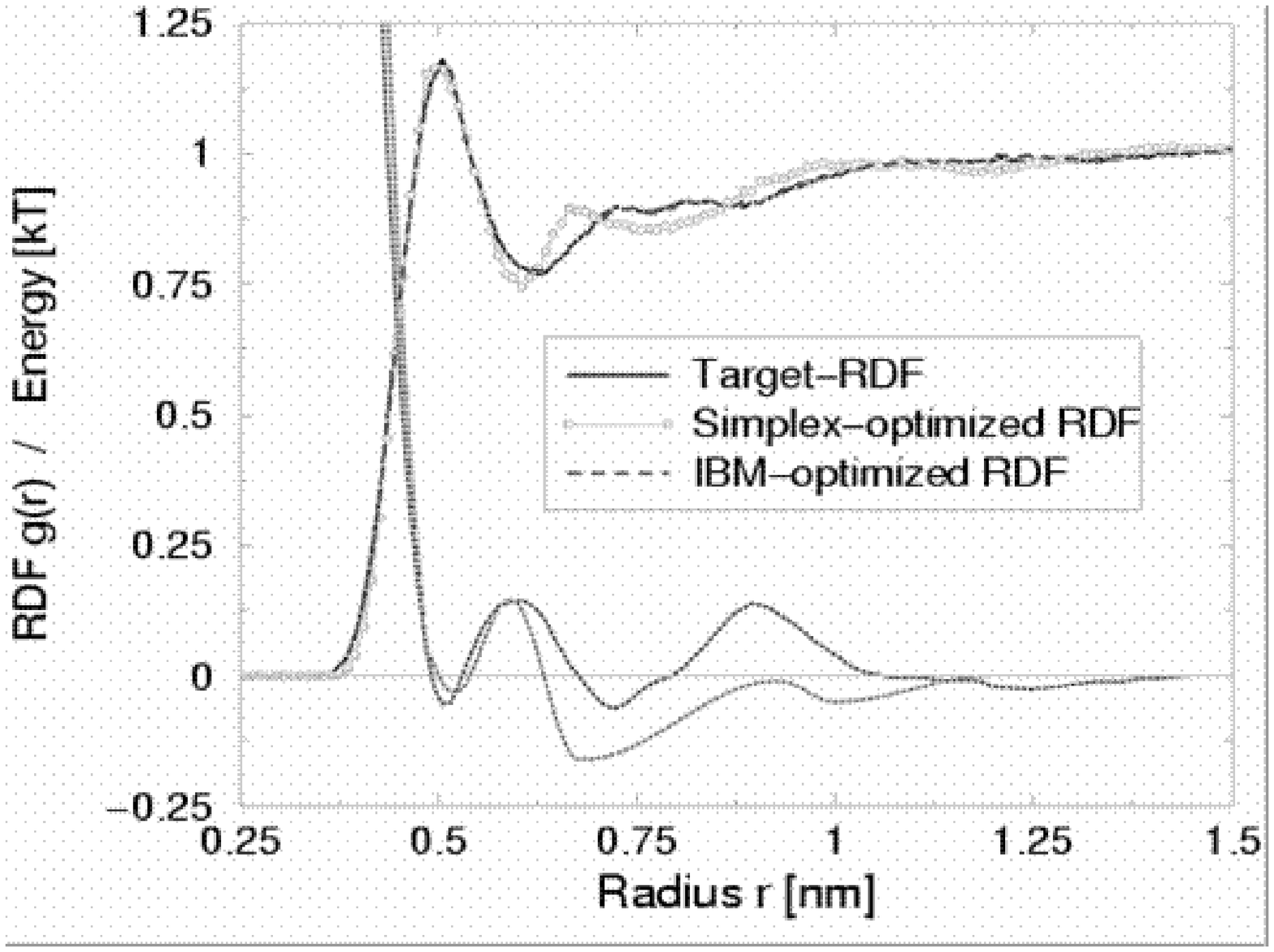}
\caption{Top: Schematic explanation of the Iterative Boltzmann
procedure. On the lower left hand side (step 2) different stages of
the potential are shown, on the upper right hand side (step 3) the
corresponding radial distribution functions are depicted. Note that
these sketches are for illustrative purposes only in order to
emphasize the influence of the iteration. Final radial distribution
functions and potentials for polyisoprene~\cite{faller03a,reith03} are
shown in the bottom part of the figure. The target function (solid
line) and the one gained by the iterative Boltzmann method (marked
IBM--optimized, dashed line) are indistinguishable.  For comparison a
simplex optimized structure (open circles) is shown. The resulting
potentials are in the lower part of that subfigure.}
\label{fig:IBM}
\end{figure}

It becomes immediately clear from this approach that the resulting
potential is numerical, as every single bin of the potential as a
function of distance is optimized independently. It is possible and
advantageous to enforce continuity by using weighted local
averages. This is important if the function to be optimized against is
relatively noisy, however, the correct way to lower the noise level 
is a longer atomistic simulation which of course can be
prohibitive. Figure~\ref{fig:IBM} illustrates the different stages of
a iterative Boltzmann procedure. In the beginning a starting potential
$V_{start}$ has to be guessed. Either we take the result from a
similar problem or we start with the potential of mean force by
Boltzmann inversion of the target function. After this initial
potential is simulated the radial distribution function is obtained
and the difference between this function and the target is
determined. This leads to a correction potential which is the
difference in free energy
\begin{equation}
\Delta V (r) = - k_B T \ln \left( \frac{g(r)}{g_{target}(r)}\right).
\end{equation}
This correction potential is added and the iteration resumes until the
difference in $g$ is deemed satisfactory.  For polyisoprene 4
iterations were necessary~\cite{reith03}. The final result is shown
at the bottom of figure~\ref{fig:IBM}.

Two alternatives to the iterative Boltzmann technique which also rely
on a physically inspired optimization of the system have been proposed
by Akkermans~\cite{akkermans00,akkermans01}. The degrees of freedom of
the polymer under study are separated into degrees of freedom of
``blobs'' and the ``bath''. The blobs play the role of the
super--atoms, the bath are all other degrees of freedom which have to be integrated
out. Only the super--atoms are taken into account. The target radial
distribution function is expanded in a basis set with the pre--factors
left for optimization
\begin{equation}
g_{target}=\sum_i\lambda_iu_i(r)
\end{equation}
The set of parameters $\lambda$ can now be viewed as dynamical
parameters and assigned a virtual mass $m(\lambda)$ and a
velocity. So one takes the route of an extended ensemble which is well
known in molecular dynamics of constant pressure and
temperature~\cite{andersen80,allen87,frenkel96}. A Lagrangian
including the $\lambda$ parameters is used and the simulation proceeds
using this extended Lagrangian
\begin{equation}
L=K(\vec{V})+K_{\lambda}(\vec{v}_{\lambda}-U(\vec{R},\lambda)-\Phi_{\lambda}(\lambda)
\end{equation}
where the $K$ stands for the kinetic energies and $U$ and $\Phi$ are
the respective potentials. Akkermans et al. showed the feasibility of
this technique by re--optimizing a Lennard--Jones potential. As the
dynamics of the $\lambda$ parameters turns out to be problematic the
same approach without the velocities can been used in a Monte Carlo
procedure.

A caveat is in order here. As all the techniques described up to now
only aim at the structure of the polymeric system it is not guaranteed
that the thermodynamic state is correctly described. This has
been pointed out by a number of
researchers~\cite{akkermans01,briels02,reith03}.  In order to avoid
such problems an inclusion of thermodynamic properties in the
optimization scheme is necessary. For the pressure in the case of the
inverted Boltzmann technique such a generalization is possible and
works as follows~\cite{reith03}. After optimizing the structure an
additional pressure correction (pc) potential of the form
\begin{equation}
\Delta V_{pc}(r)=A_{pc}\left(1-\frac{r}{r_{cut}}\right)
\end{equation}
is added, where $A$ is negative if the pressure is too high and
positive if it is too low. The rationale behind this choice
is to have a constant force in addition to the force from the
structural potential which leads to a constant shift in pressure. With
such an additional potential the radial distribution function does not
deteriorate strongly and a re--optimization is possible.  Reith et
al. showed that indeed this pressure correction solved their initial
problem of an unphysically high pressure~\cite{reith03}.
\section{Dynamic Mapping}
Simulations in atomistic detail regularly utilize a time--step of 1
femtosecond. This time--step has to be about an order of magnitude
shorter than the fastest characteristic time of the system. As
customarily the bond lengths are fixed using techniques like {\sc
Shake}~\cite{ryckaert77,mplathe91} or {\sc
Rattle}~\cite{andersen83,allen87} the fastest time--scales in
atomistic molecular dynamics are bond vibrations on the order of tens
of femtoseconds. With a reasonable use of computer resources one can
then reach into the nano--second time--range. This is long enough to
compare to segmental dynamics in NMR
experiments~\cite{faller01a,budzien02} but not long enough to compare
to large time--scale experiments.

The techniques to map the statics of polymers which have been
described above lead inherently to larger time--scales as the
fastest inherent degrees of freedom are now motions of super--atoms
of the size of monomers. If dynamic
investigations are desired one has to find a correct mapping of
the time--scales of the atomistic simulation to the meso--scale.
Otherwise dynamic experimental comparisons are impossible.
\subsection{Mapping by chain diffusion}
An obvious candidate for calibrating the time--scale is the chain
diffusion coefficient. At large enough times any polymer chain in a
melt will end up in diffusive motion as soon as all internal degrees
of freedom are relaxed. This diffusion can be used to determine the
time--scale as long as an independent mapping of the length scale is
achieved. The static mapping determines the length scale; an obvious
choice is the size of the monomer or the distance between super--atoms
along the chain to obtain a length scale for the coarse--grained
simulation~\cite{faller02a}. If both simulations, the atomistic and
the coarse grained can be fully equilibrated in the sense that free
diffusion of the whole chain is observed the two diffusion
coefficients can be equated and the time--scale is fixed. In most
cases a full free diffusion of the atomistic chain can not be reached
in reasonable computer time. This is especially the case when the
coarse--grained simulation should be used as a means to efficiently
equilibrate the structure from which atomistic simulations will be
started.

Nonetheless this technique can be successful. In the case of
10--mers of polyisoprene at 413K a dynamic mapping between a fully
atomistic and a very simple coarse grained model is
possible~\cite{faller01a,faller02a}. Only chain stiffness was used to perform
the mapping. The local chain reorientation in both simulations was
the same after the time--scales had been determined by the
diffusion coefficient. However, the decay times of the Rouse modes
were not equal which showed that the mapping by stiffness alone
was too simplistic.
\subsection{Mapping through segmental correlation times or Rouse model}
It is often easier to use shorter, local, time scales to map the
atomistic to the coarse--grained length scale. 
This allows a mapping also if the atomistic simulation
cannot be simulated into free diffusion. Even if free diffusion
can be reached the statistical uncertainty of large time scales is
often so large that a shorter time scale is a better
choice for the mapping. Candidates for shorter time--scales are 
decay times of higher Rouse modes.  Even if the Rouse model is not 
a perfect description of the system under study such 
a mapping remains meaningful. In that case this time still corresponds to a 
well defined relaxation time of a chain segment.

If such a chain segment consists in the extreme case of only one
monomer we end up with the segmental relaxation time or equivalently
the reorientation on the monomer scale. This time--scale is very useful
for dynamic mapping as it can be compared the time--scales in NMR
experiments~\cite{doxastakis03}.
\subsection{Direct Mapping of the Lennard--Jones time}
A completely different idea which is independent of the atomistic
simulation is the mapping of the Lennard--Jones time to real time.
If one applies the standard Lennard--Jones units where we measure
lengths in $\sigma$, the particle diameter, energies in $\epsilon$
the depth of the Lennard--Jones potential, and masses in $m$ the
monomer mass, naturally a timescale appears which is
conventionally called the Lennard--Jones
time~\cite{allen87,frenkel96}.
\begin{equation}
\tau=\sigma\sqrt{\frac{m}{\epsilon}}
\end{equation}
This time--scale can be used to perform the mapping to the
real time--scale~\cite{reith01b,faller03a}.

Using the polyisoprene models of ref.~\cite{faller01a} (atomistic at
$T=413\text{K}$) and ref.~\cite{reith03,faller03a} (meso--scale) we
get the following differences in the center--of--mass diffusion
coefficient for a atomistic 10--mer: The Lennard--Jones time leads to
$D_{com}=16\times10^{-6}\text{cm}^2/\text{s}$~\cite{faller03a}. If we
map the diffusion coefficient directly $D$ is obviously the atomistic
result of $D=4.24\times10^{-6}\text{cm}^2/\text{s}$. This result was
actually obtained by matching the center--of--mass motion of two
different models and fitting the large scale motion of the coarser
model. This was necessary as even at 413~K the simulation does not move
the atomistic 10--mers into free diffusion~\cite{faller01a}. Recently
for $cis$--polyisoprene a united atom model could be brought into free
diffusion~\cite{doxastakis03}. In this case results for 8--mers
($D=14\times10^{-6}\text{cm}^2/\text{s}$) have been reported which are
close to the results for the different $trans$--PI--models. This
indicates that the different mappings are not far from each other but
a uncertainty of the order of 2--5 has to be taken into account. For
polyisoprene this mapping actually gives a reasonable description of
the experimental diffusion coefficient~\cite{doxastakis03}.
\section{Automatic Optimization -- In Coarse Graining and Elsewhere}
Polymer coarse--graining is by no means the only or even the first
area of computer simulations where automatic optimization techniques
are applied.  Already in the 70s Torrie and
Valleau~\cite{torrie74,torrie77} proposed a Monte Carlo technique to
simplify simulations in complex energy landscapes which can easily be
implemented fully
automatically~\cite{beutler94,roux94,bartels97}. This so--called
umbrella sampling bases on the idea that any bias in a Monte Carlo
simulation can be used as long as it is taken into account in the
analysis. For sampling reasons a uniform coverage of the interesting
energy area is of advantage as in that case the system does not get
trapped in any configuration but samples the whole configuration space
readily. Umbrella sampling has been recently combined with parallel
tempering to get a fully automatic multicanonical parallel tempering
scheme~\cite{faller02c}. An idea similar in spirit to umbrella
sampling is density of states Monte Carlo which even in its very first
implementation~\cite{wang01a,wang01b} was a completely automatic
procedure. It abandons the detailed balance criterion of Monte Carlo
in its early stages of sampling in order to get a better automatic
optimization. This technique has since been generalized and improved
in a number of
ways~\cite{calvo02,yan02,kim02,faller03b,shell03,troyer03,yan03}. All
have in common that they aim at an automatic calculation of the free
energy and in that sense the iterative Boltzmann method discussed
above is only a special case of this much broader class of
techniques. It may be worthwhile to think about method transfer
between the Monte Carlo calculations of the partition function as
aimed by umbrella sampling or density of states Monte Carlo and
polymer coarse graining.

Conclusively one can say that the recent efforts in automatic polymer
coarse--graining have led to a a number of very efficient and
systematic techniques to map atomistic models onto meso--scale
models. Especially, the thermodynamically inspired iterative
Boltzmann technique is fast and reliable for a number of systems.
The main drawback is still the dependence on the single state
point. In the transition from the atomistic to the coarse--grained
scale we gain a lot of efficiency but loose the generality of the
atomistic model as the coarse--grained model is optimized to the
atomistic simulation at a defined state point. Especially in an
effort to generalize the coarse graining to polymer mixtures this
problem becomes apparent~\cite{sun03p}

The state of the art in dynamic mapping is much less clear than
the structural optimization. As the optimized force--fields up to
now aim exclusively at the structural or thermodynamic properties
the dynamic mapping is an ad hoc step which may or may not be
successful. This is especially true if solutions are to be mapped
as the idea of coarse--graining is to get rid of the solvent.
However, the solvent has a marjed effect on the dynamics which in the
coarser simulations without solvent is not present. To overcome this problem and 
include the dynamic effects of the solvent without explicit solvent
lattice--Boltzmann simulations may be the way to
go~\cite{ahlrichs99}. In the case of melt simulations the solvent
effects are not the problem but the resulting force--fields are up
to now not able to get all the characteristic times correct
at the same time so that a lot of work remains to be done.
\section*{Acknowledgments}
Many fruitful discussions with Markus Deserno, Juan de Pablo, Kurt
Kremer, Florian M\"uller--Plathe, Hendrik Meyer, Dirk Reith, and Doros
Theodorou are gratefully acknowledged. I especially want to thank Qi Sun 
for some analysis of the {\it cis}--PI system.
\bibliography{standard}
\bibliographystyle{polymer}


\end{document}